\documentclass[aps,10pt,groupedaddress,onecolumn,superscriptaddress,prl]{revtex4-1}

\usepackage[utf8]{inputenc}
\usepackage{graphicx}
\usepackage{dcolumn}
\usepackage{bm}
\usepackage{paralist}

\usepackage{booktabs}
\usepackage{glossaries}
\usepackage{multirow}

\newacronym{afc}{AFC}{Automatic Fare Collection}
\newacronym{rods}{RODS}{Rolling Origin and Destination Survey}
\newacronym{tfl}{TfL}{Transport for London}
\newacronym{od}{OD}{Origin-Destination}
\newacronym{pt}{PT}{Public Transit}
\newacronym{si}{SI}{Supplementary Information}
\newacronym{gmm}{GMM}{Gaussian Mixture Model}
\newacronym{cbd}{CBD}{Central Business District}
\newacronym{tod}{TOD}{Transit-Oriented Development}

\begin{document}

\title{Extracting Spatiotemporal Demand for Public Transit from Mobility Data}

\author{Trivik Verma* \thanks{Correspondence and requests for materials should be addressed to TV (t.verma@tudelft.nl)}}
\affiliation{Faculty of Technology, Policy and Management, Delft University of Technology, 2628BX Delft, the Netherlands}

\author{Mikhail Sirenko}
\affiliation{Faculty of Technology, Policy and Management, Delft University of Technology, 2628BX Delft, the Netherlands}

\author{Itto Kornecki}
\affiliation{ETH Z\"urich, Universit{\"a}tstrasse 16, 8092 Z\"urich, Switzerland}

\author{Scott Cunningham}
\affiliation{Faculty of Humanities and Social Science, University of Strathclyde, 18 Richmond Street, Glasgow G1 1XQ, Scotland, United Kingdom}

\author{Nuno A. M. Ara\'ujo}
\affiliation{Departamento de F\'isica, Faculdade de Ci\^encias, Universidade de Lisboa, P-1749-016 Lisboa, Portugal}

\affiliation{Centro de F\'isica Te\'orica e Computacional, Faculdade de Ci\^encias, Universidade de Lisboa, 1749-016 Lisboa, Portugal}

\begin{abstract}

With people constantly migrating to different urban areas, our mobility needs for work, services and leisure are transforming rapidly. The changing urban demographics pose several challenges for the efficient management of transit services. To forecast transit demand, planners often resort to sociological investigations or modelling that are either difficult to obtain, inaccurate or outdated. How can we then estimate the variegated demand for mobility? We propose a simple method to identify the spatiotemporal demand for public transit in a city. Using a Gaussian mixture model, we decompose empirical ridership data into a set of temporal demand profiles representative of ridership over any given day. A case of $\approx 4.6$ million daily transit traces from the Greater London region reveals distinct demand profiles. We find that a weighted mixture of these profiles can generate any station traffic remarkably well, uncovering spatially concentric clusters of mobility needs. Our method of analysing the spatiotemporal geography of a city can be extended to other urban regions with different modes of public transit.
\end{abstract}
\maketitle

\section*{Introduction}

According to the UN, every year about $1$ billion people migrate \cite{beaverstock_lending_1996,darling_forced_2017} to different urban areas around the world \cite{united_nations_department_of_economic_and_social_affairs_desa_crossnational_nodate}. With accelerated migration rates, urban areas expand and citizen needs grow considerably \cite{alonso_theory_1960}. \gls{pt} networks play a significant \cite{rodrigue_geography_2013} and challenging role \cite{zhang_network_2019} in serving citizens' business, industrial, social, cultural, educational and recreational needs \cite{ceder_public_2016}. A prevalent urban policy of \gls{tod} promotes mixed-use of urban areas by encouraging people to live near clusters of amenities and services, built around transit stations. Such development enhances accessibility of the city for a wider group of citizens \cite{tomer_missed_2011}, improves pedestrian access \cite{dittmar_new_2012}, and reduces pollution and congestion by discouraging reliance on private vehicles \cite{calthorpe_next_1993}. Though cities welcome innovation in infrastructure services necessary for sustaining growing populations \cite{Bettencourt2007}, improved connectivity also threatens to create larger, more sprawling urban areas \cite{henig_gentrification_1980,dawkins_transit-induced_2016}.

Due to the changing transit needs \cite{smith_no_2013} and developments in urban land-use, the planning and management of transit services pose a huge challenge \cite{byrne_complexity_2003}. Transportation demand analysis heavily depends on census-based population statistics and usage estimates from transit authorities \cite{ceder_public_2016}. Some researchers emphasise qualitative methods to estimate transport demand, involving direct observation of urban populations \cite{Raudenbush1999,taylor_nature_2009,gutierrez_transit_2011}. Studies that focus on revealing macroscopic urban structures \cite{Anas1998,Barthelemy2011,burgess_growth_2008} develop aggregated \gls{od} trajectories of people using mobile phone \cite{noulas_exploiting_2013,calabrese_estimating_2011,Louail2014,Louail2015} and twitter data \cite{mcneill_estimating_2017}. While there is evidence that urban mobility patterns are reproducible using aggregated statistics of populations \cite{gonzalez_understanding_2008}, this data only accounts for $2-5\%$ of the entire population \cite{Louail2015}. What is more, recent work suggests methods based on incomplete statistical data underestimate important trips, especially in larger cities \cite{camargo_diagnosing_2019}. This general framework of estimating a snapshot of transit demand and adapting future supply to it relies on several parameters, uses incomplete data, and is difficult to compare through time.

Over the past decade, digital services like \gls{afc} have been introduced into transit networks worldwide. There is important literature on flow estimation that extracts detailed and complete \gls{od} trajectories \cite{Park2008,roth_structure_2011,zhong_detecting_2014,long_combining_2015} from \gls{afc} data. Using complete data, these studies provide aggregated instances of mobility flows between large hotspots in a city, mainly focusing on recovering morphological characteristics of the urban structure \cite{Anas1998,Barthelemy2011}. But the spatiotemporal nature of mobility lacks a strong empirical foundation. Identifying the daily demand for mobility across time and space can naturally structure the efficient development and management of \gls{pt} services catering to our complex mobility needs \cite{domenico_navigability_2014}.

We propose a simple method to estimate the varying demand for transportation in time using anonymous, privacy-preserving, complete and freely available entry-only ridership data. Our case uses the data of $\approx 4.6$ million daily commutes in the \gls{tfl} services in the Greater London Underground network. We find that the daily traffic through this \gls{pt} network is a mixture of \emph{six} demand profiles. Using the weights of these profiles, our model is able to reproduce individual network station traffic throughout the day and classify stations into \emph{six} categories. Upon mapping these categories, we show how these profiles can identify the spatial distribution of varying mobility demand for the \gls{pt} infrastructure. We discuss how the temporal nature of complex urban demand reveals the spatial structure of the city consisting of central \cite{murphy_central_2017}, polycentric \cite{roth_structure_2011,Louf} and concentric \cite{burgess_growth_2008,hoyt_structure_1939} zones of development. We expect our method to be useful for data-driven demand analysis of \gls{pt} infrastructures in any region in the world where complete \gls{od} trajectories are not accessible or are only available through proprietary sources.

\section*{Results}
\textbf{Beyond traditional transportation demand analysis.} To measure the changing demand for \gls{pt} we analyse the passenger entrance counts, $P_i(t)$, entering a station $i$ at time interval $t$ ($\forall t \in {1, 2, ..., m}$) where each interval is a $15$-minute observation window in which the data is collected with $m$ intervals per day. The variable $P_i$ is a proxy for the ridership behaviour, an indication of the usage of every station at different times in a day. We use $\approx 4.6$ million geolocated observations of daily \gls{tfl} passengers. Our dataset consists of $264$ stations of the Greater London region spanning $24$ hours in a day (see the Methods section for details).

\begin{figure}
\includegraphics[width=0.55\textwidth]{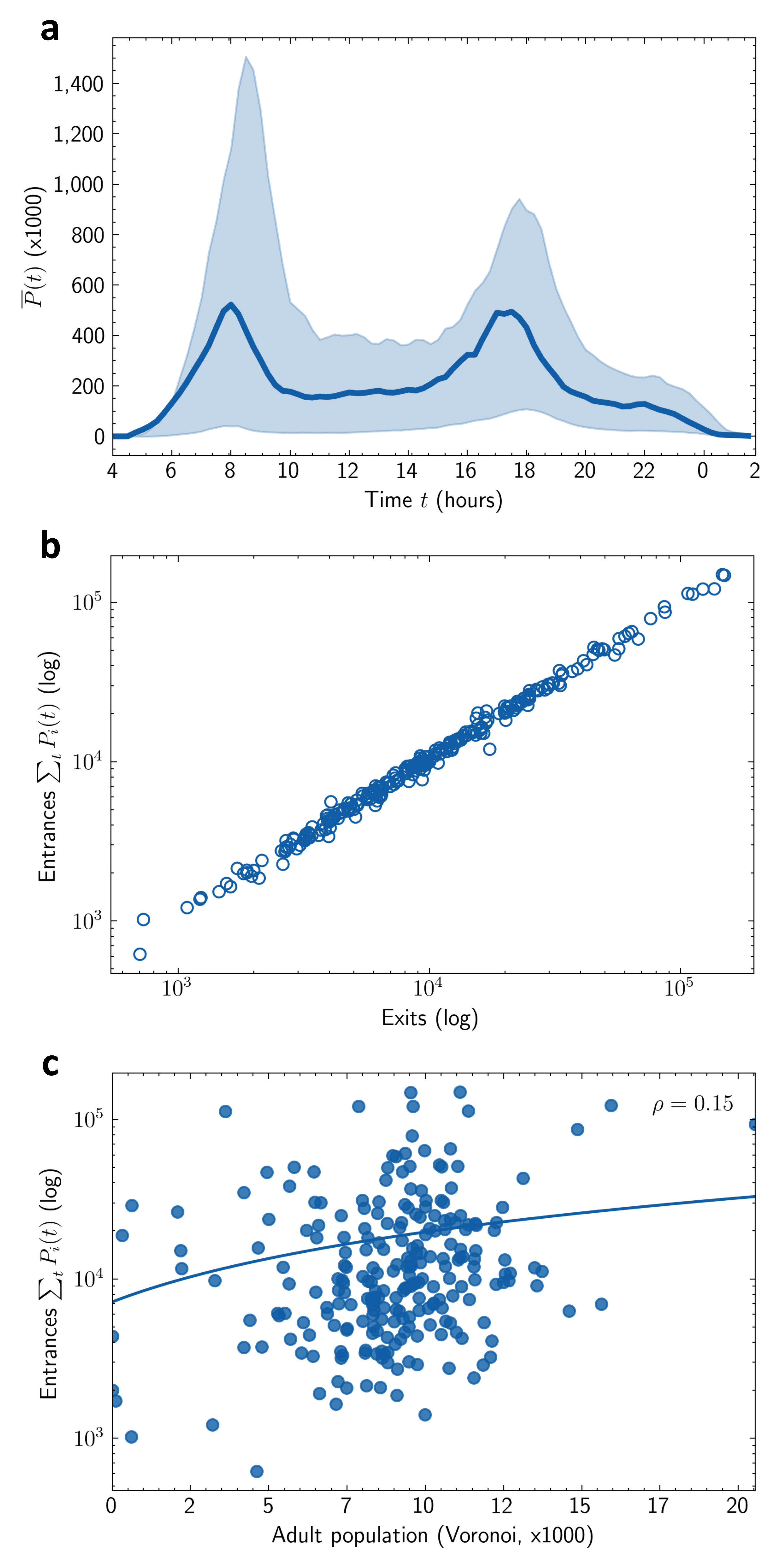}
\caption{\textbf{Describing the \gls{tfl} data of station traffic over a representative day. a} Average Passenger counts $\overline{P}(t)$ for the \gls{pt} system showing entrances for every $15$-minute intervals. The confidence intervals show the quantile interval ($10-90\%$ of data) and the blue line is the average traffic for each cluster \textbf{b} Relationship between the total entry vs exit counts for every station in the system. \textbf{c} Relationship between population and ridership (entry counts) for Voronoi cells that are attributed to every station (see Supplementary Information for estimating population counts for station zones). The number of stations in this figure are less than the total number in the dataset because some stations are outside Greater London for which population estimates were not available}
\label{fig:fig1}
\end{figure}

Figure \ref{fig:fig1}a represents the average demand for the \gls{pt} system across all stations in the network throughout the day with the quantile interval showing variations across all stations in the city (see Supplementary Figure 1 for an overview of the entire system traffic). As it is evident from Fig. \ref{fig:fig1}b, aggregate station ridership is symmetric: data from a day shows excellent correlation between the total number of entrances and exits for every station. To understand how station traffic is related to population statistics, we visualise the relationship between the number of total station entrances on a given weekday, $\sum_t P_i(t)$, and the working adult population of every zone associated with the station (see Supplementary Note 1 for details on estimating population sizes for station zones). Figure \ref{fig:fig1}c shows a very weak relationship between the number of people residing in a zone and the entrances at the corresponding station. The weak correlation suggests population statistics are not a good proxy for demand. If people are not using the station closest to them, that may be due to some stations not fulfilling the potential for accessing opportunities in a city. Even though aggregate counts of entrances and exits are matched well (Fig. \ref{fig:fig1}b), an indication of the complexity of the urban demand for mobility can be witnessed in asymmetric correlations between trips made in opposite directions illustrating the increasing use of stations for other activities than work and residential (see Supplementary Note 2 for details). Considering this evidence, census-based populations surveys are not accurate enough to calibrate gravitation models \cite{batty_urban_1976} for estimating complex city flows and are often found to underestimate importance of regular home-work trips \cite{camargo_diagnosing_2019}. Thus, we focus on identifying distinct demand profiles across a day.

\begin{figure}
\includegraphics[width=0.9\textwidth]{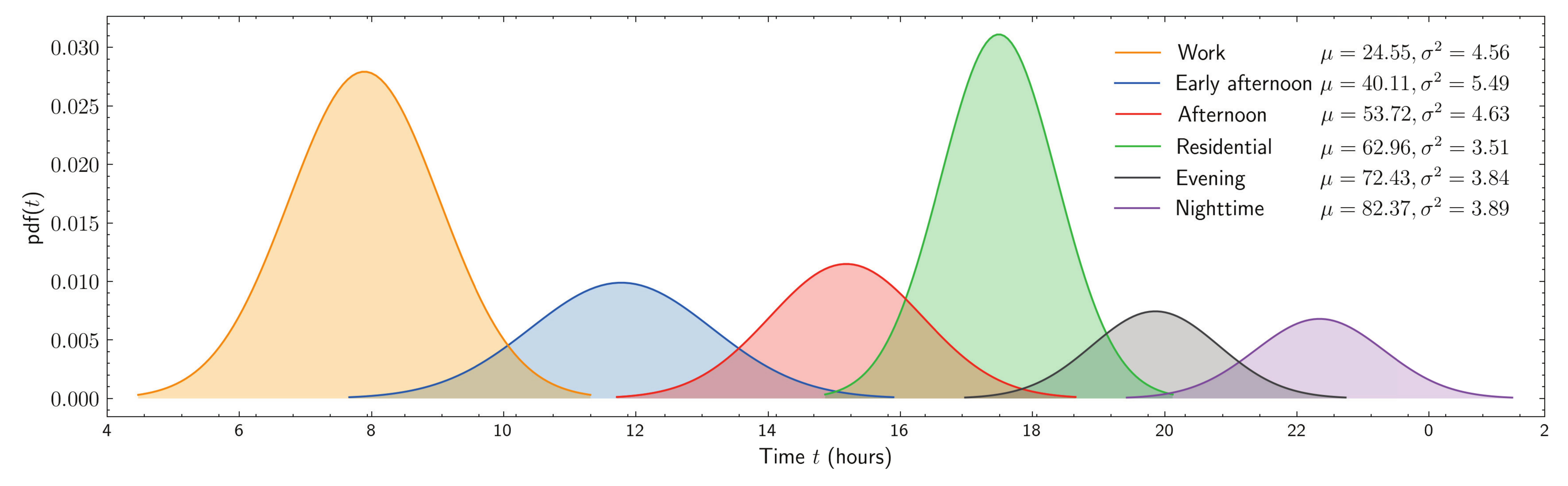}
\caption{\textbf{Temporal demand profiles over the representative day.} Each normal distribution refers to a particular kind of traffic in time and the density estimates $pdf(t)$ measures the likelihood that an individual entering the system belongs to a certain demand profile. $\sum_k pdf(t) = 1$.}
\label{fig:fig2}
\end{figure}

\textbf{Multiple demand profiles for public transit.}
The time series of entrances at stations represents an aggregation of many different commuting patterns. To interpret the variability in demand, we formulate a simple \gls{gmm} (see the Methods section for details) to identify temporal patterns in the time series of $\sum_i P_i(t)$. The model represents the cumulative transportation demand as normal distributions with varying mean ($\mu_k$), variance ($\sigma_k$) and mixture weights ($\phi_k$) for each demand profile $C_k$. This choice is in part justified by the multimodal distribution of traffic over the \gls{pt} network across time (Fig. \ref{fig:fig1}a). We identify \emph{six} demand profiles that are all normally distributed in time (see Supplementary Figure 6 for variation in demand profiles). Figure \ref{fig:fig2} illustrates the distribution of the characteristic mixtures of subpopulations in the data, each with varying parameters ($\mu, \; \sigma \; and \; \phi$: see Supplementary Table 1 for details.)

Each mixture represents one transportation demand profile \cite{smith_no_2013,noauthor_london_2018}:
\begin{enumerate}
\item Work (W): Morning trips for work;
\item Early Afternoon (LM): Late Workers, tourists, shoppers and miscellaneous activities;
\item Afternoon (A): School and lunchtime traffic, flexible workers, tourists, and shoppers;
\item Residential (R): Evening trips returning from work;
\item Evening (E): Late workers returning home and dinnertime traffic;
\item Nighttime (N): Service industry (restaurant, bars, healthcare) workers, and traffic from entertainment districts.
\end{enumerate}
The profiles mentioned above are generalisations of traffic but there is mixed usage throughout the day: tourists travel at all times and night workers come home in the morning as well. In this work, we do not categorise individual traces into any types.

\textbf{Clustering stations by demand profiles.}
The demand profiles we identify using the \gls{gmm} are expressed using three parameters, a mixture component weight, mean and variance. Using this set of values we reconstruct individual station traffic for every station by representing the demand of a station as a linear combination of different types and classifying the station based on the relative value of the weights and the three estimated parameters of the \gls{gmm}. The estimated probability density reveals the subscription of each station to every mixture (see the Methods section for details on generating station traffic and clustering).

\begin{figure}
\includegraphics[width=0.9\textwidth]{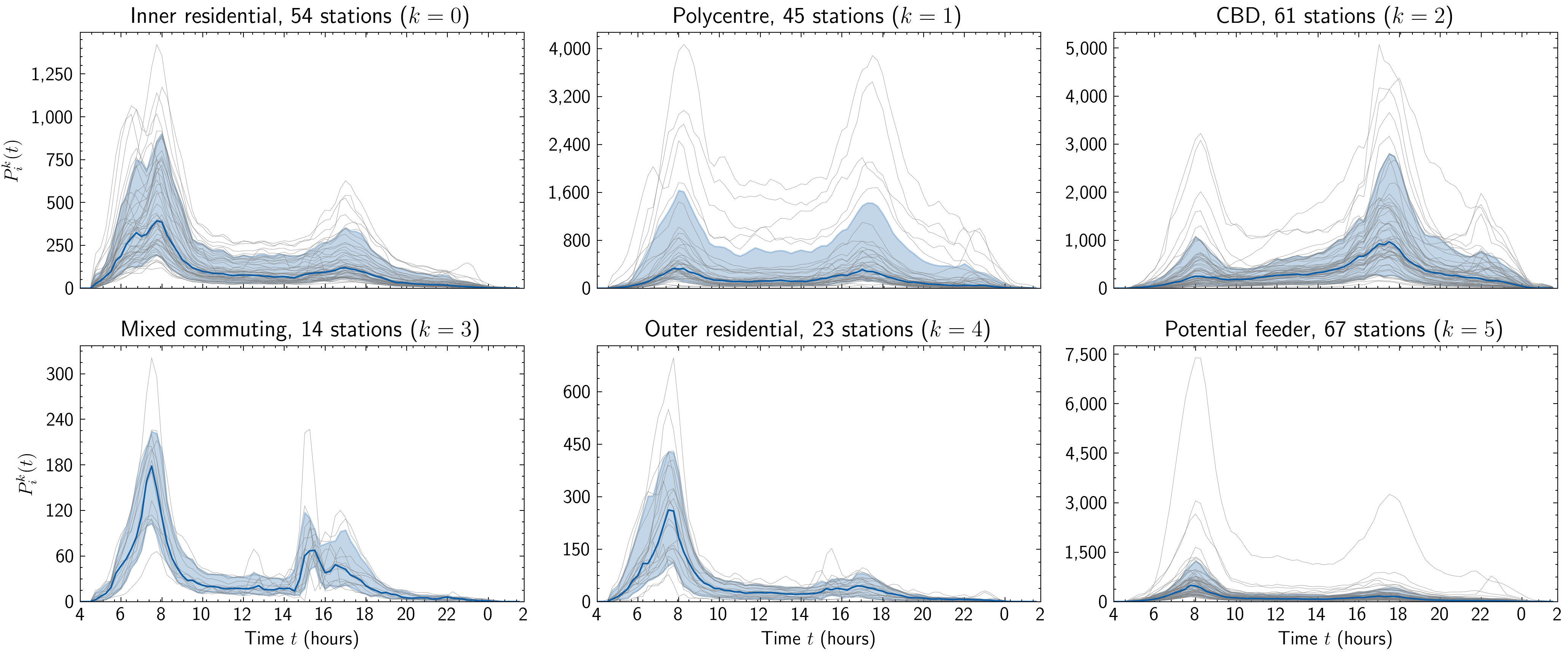}
\caption{\textbf{Clusters of generated station traffic.} \emph{Six} clusters of station traffic showing different peculiar patterns of station use. Each sub plot is the time series of generated station traffic $P_i^k(t)$. The confidence intervals show the quantile interval ($10-90\%$ of data) and the blue line is the average traffic for each cluster. The thin grey lines show all station traffic belonging to a cluster. All $264$ stations of the Greater London region are shown here.}
\label{fig:fig3}
\end{figure}

We analyse the different clusters of stations in Fig. \ref{fig:fig3}. Each cluster has a particular shape of average daily traffic entering the station:
\begin{enumerate}
\item \emph{\gls{cbd}} stations show a higher number of entrances in the evening compared to the rest of the day. People usually move to the district for using services and business all day and return home at night;
\item \emph{Polycentre} stations witness similar amount of workbound traffic in the morning and residential traffic in the afternoon, and a high amount of activity throughout the day compared to other clusters. These are large secondary hubs \cite{Louail2015} that have mixed use for residences, workplaces and services;
\item \emph{Potential feeder} stations that are in a zone of transition \cite{hoyt_structure_1939} with changing land-use from a compact and busy \gls{cbd} to wider residential regions with self-sufficient services. A peak in workbound traffic and a declining residential traffic pattern suggests middle-class housing workers residing in this region possibly feed into the city for work;
\item \emph{Inner residential} stations may be serving working-class groups but are further away from the \gls{cbd} (see Supplementary Figure 7). These stations differ from the feeders because of a striking peak in the early morning traffic just before the workbound ridership peaks;
\item \emph{Outer residential} stations have lesser evening traffic and are much further from the \gls{cbd} compared to inner residential stations. The lower evening volumes point to the region's greater residential nature;
\item Finally, \emph{commuter} stations have a mix of suburban or satellite traffic in the morning and residential traffic in the evening, pointing to some work locations in the vicinity as is expected from clusters of suburban areas.
\end{enumerate}

\begin{figure}
\includegraphics[width=0.9\textwidth]{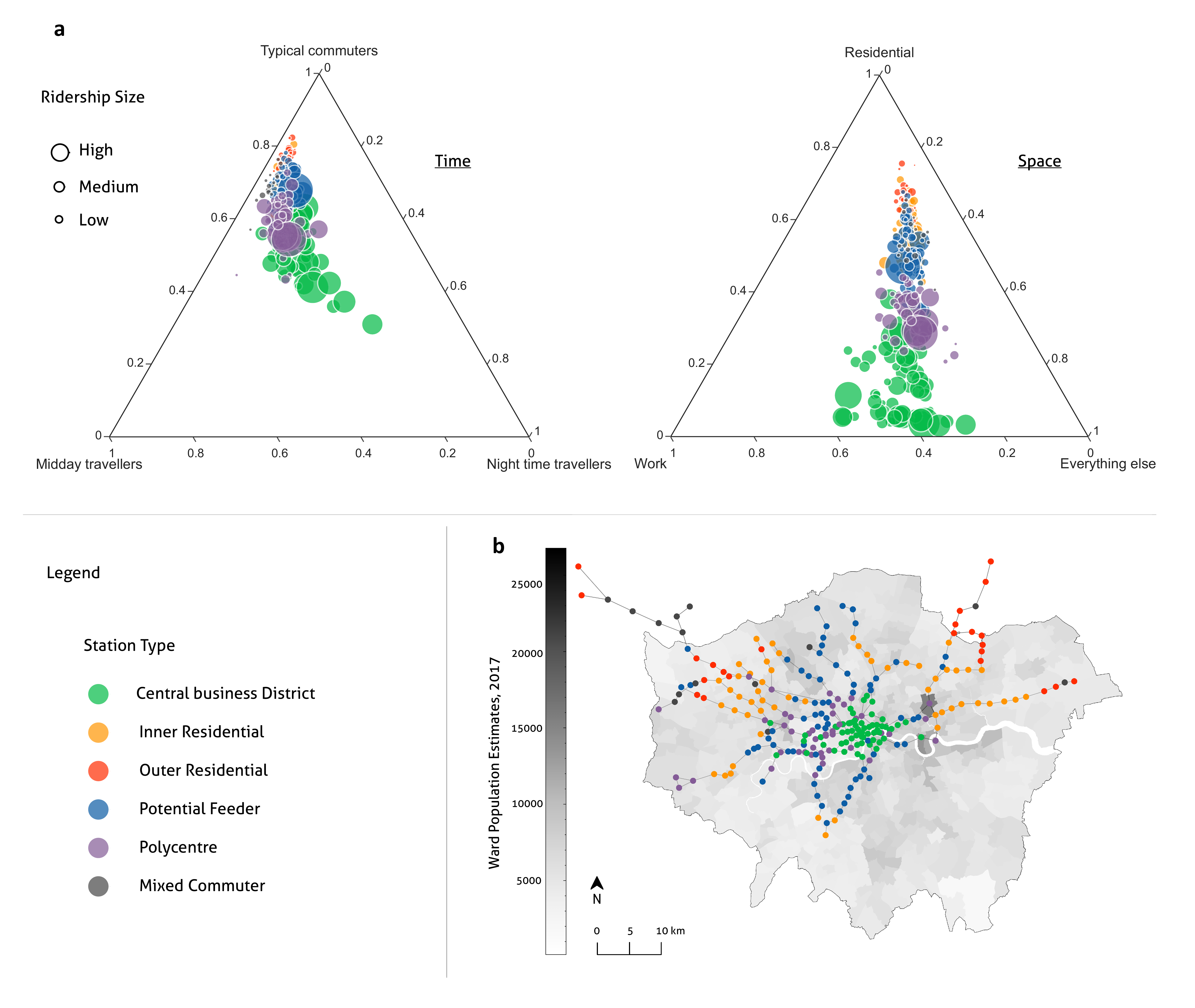}
\caption{\textbf{The spatiotemporal geography of the London public transit system. a} The two ternary plots show scenarios of time and space where demand profiles are aggregated by time: typical (W and R), midday (A and EA) and after-work (E and N), and space: work (W), residential (R) and other (A, EA, E and N) \textbf{b} The station clusters mapped onto the Greater London region with locations as the coordinates of the underground service. The gradient in grey is a population map showing the adult working population of the Wards from a $2017$ intercensal estimate.}
\label{fig:fig4}
\end{figure}

Figure \ref{fig:fig4}a reports two spatiotemporal scenarios. In the first one (left frame), we show how station clusters are distributed among typical (W and R), midday (LM and N) and nighttime (EN and LN) demand profiles. Note that the \gls{cbd} is skewing the distribution toward nighttime traffic, possibly because entertainment centres are located close to business districts. The plots also reveal that the \gls{pt} system in London is used much less in the midday hours in comparison to the regular home-work traffic at other times of the day. The second scenario (Fig. \ref{fig:fig4}a - right frame) shows a prevalence of work (W) and other (LM, N, EN and LN) traffic profiles than residential (R) in the \gls{cbd} stations. It is possible that many residents in the \gls{cbd} region who have improved access to bus services and therefore do not use underground transit. Additionally, these residents may be affluent and own a car, live within walking / biking distance of their place of work or are older, non-working citizens. Also, there are more residential and typical stations in the city and these stations have less traffic than the business district. This finding is consistent with the “many-to-few” characterisation of a typical city \cite{roth_structure_2011}, where many residential areas feed a small number of polycentres and the \gls{cbd}.

Visualising the station clusters in space in Fig. \ref{fig:fig4}b, mapped onto the public transport system, reveals how the city is organised in concentric circles as argued and proposed by Burgess \emph{et al.} \cite{burgess_growth_2008}. The spatial distribution of clusters describes a complex urban structure wherein public transit links the \gls{cbd} to outer residential spaces through distinct linkage patterns, revealing a monocentric structure with respect to the primary means of underground transportation. Possibly, the users enter the system from the periphery and advance to the \gls{cbd} for work or other activities. This indicates that most residential areas are spread out at the outskirts of the city, while the stations close to work centres are clustered together in the \gls{cbd}. The map thus reveals that there is a dense urban core that is the City of London, surrounded by a residential periphery. This is supported by measuring the average travelling distance between the stations, where we find that the stations in work districts are much closer together than ones in residential districts (see Supplementary Figure 8 for details). Our analysis confirms both concentric \cite{burgess_growth_2008} and polycentric nature \cite{roth_structure_2011} of the city. Polycentres also include some National Rail connections, such as Waterloo, Brixton, and Stratford stations. Though some of these outliers appear as important polycentric hubs for the city (for example, Waterloo), the stations themselves may not have many residents in the vicinity using the system. They are important connections which collect residents coming from other cities via the national train network.

\begin{figure}
\includegraphics[width=0.9\textwidth]{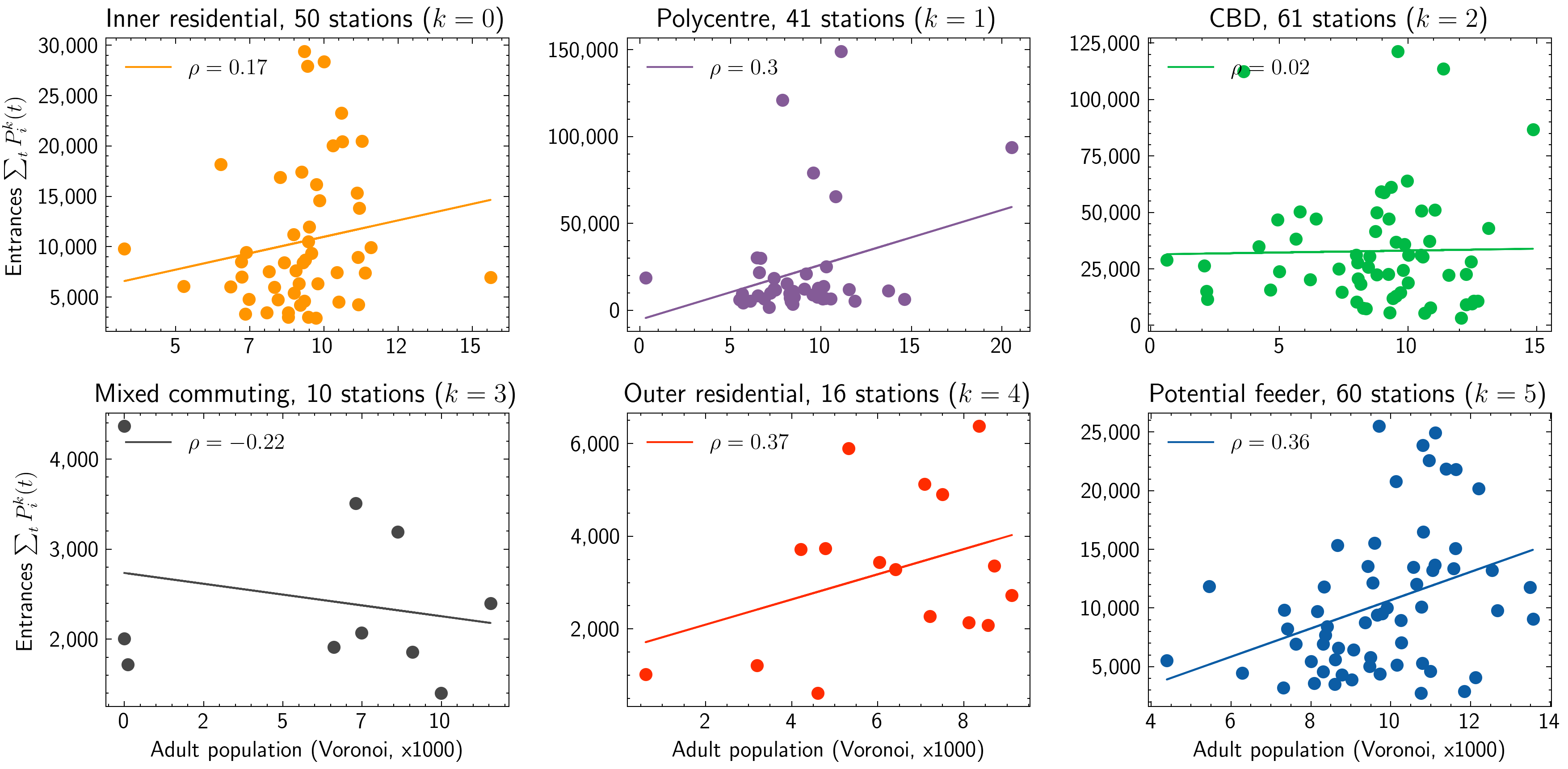}
\caption{\textbf{Cluster specific scatter-plot representation of ridership versus adult working population.} Relationship between population and ridership (entry counts) for Voronoi cells that are attributed to every station (see Supplementary Note 1 for estimating population counts for station zones). The number of stations in this figure are less than the total number in the dataset because some stations are outside Greater London for which population estimates were not available. Only $238$ sations are shown in this plot, as the remaining $26$ stations in the dataset are outside the Greater London region for which we cannot estimate population sizes and hence Voronoi tessellations.}
\label{fig:fig5}
\end{figure}

To understand the complex nature of urban flows we disentangled station usage into various clusters. Next, we individually examine the correlations between ridership and the adult working population of each zone separated by station clusters (recall Fig. \ref{fig:fig1}c for this discussion). Figure \ref{fig:fig5} illustrates that there is a stronger correlation between the ridership at a station and the population within the station’s zone for outer-\emph{residential}, inner-\emph{residential} and polycentric clusters. This is intuitive: the number of people taking the train in the morning from a residential station is proportional to the number of people living within the proximity of that station where other activity is also minimal (see Supplementary Note 1 for a discussion on accessibility analysis). Polycentres by definition are secondary hubs that attract people owing to their business/services/residential mix. As expected, the relationship for other clusters, the \gls{cbd}, mixed commuting and feeder stations, is weak and prone to outliers.

In addition to revealing the \gls{cbd} of London, our method can reveal other commercial areas in the city. For example, we observe White City Station as being commercial, which corresponds to the concentration of several large businesses, including the British Broadcasting Corporation (BBC) offices and the Westfield Mall, the largest mall in Europe. Since our method is normalised for traffic volume, it also reveals areas which, though relatively low in ridership, are heavily business-oriented. For example, there is a strong work-like ridership pattern in Canary Wharf Station, corresponding to the large financial district in the Isle of Dogs area. Thus, our modelling has the potential to analyse mixed-use of stations which has far-reaching implications in improving services (see Supplementary Note 3 for a use case explanation).

\section*{Discussion}
Digitisation has enabled an unprecedented amount of anonymous and location-based transit ridership data that is both complete (representative of the entire population using the service) and easily tractable. Our method leverages such information to extract mobility demand profiles across a city over the course of the day. It is independent of trajectories of individuals, thus preserving their privacy. Upon clustering the station traffic generated using these demand profiles, we are able to extract significant information about urban structures of a city. We have applied this method to the Greater London region using a dataset consisting of $\approx 4.6$ million transit traces. The empirical results show three key findings.

First, the aggregate usage of a public transit network can be decomposed into distinct temporal demand profiles that represent various classes of daily ridership for work, services, leisure and other combinations of use \cite{ahas_daily_2010,gordon_distribution_1986,lenormand_comparing_2015}. Second, stations clustered by their traffic patterns suggest that there are concentric zones of development in Greater London, identifying polycentres, entertainment and tourist locations, residential and highly specialised business districts. Third, larger station show mixed-use demand and stations farther away from the centre of the city are likely to exhibit a prevalent residential ridership. While there is evidence of declining populations and traffic from the central business district \cite{tobler_computer_1970}, there are rhythms to human activity \cite{smith_no_2013} and matching the various demands will lead to efficient transport utilisation across the entire network.

Transportation demand analysis in large metropolitan areas is an important problem that is relevant for urban planning \cite{byrne_complexity_2003}. Researchers either investigate detailed sociological data \cite{Raudenbush1999,taylor_nature_2009,gutierrez_transit_2011} or extract macroscopic urban structures \cite{noulas_exploiting_2013,calabrese_estimating_2011,Louail2014,Louail2015} as proxy for demand patterns. However, as transit needs evolve, intra-urban mobility structures \cite{Anas1998,Barthelemy2011,burgess_growth_2008} also transform. Incomplete data \cite{Louail2015} from digital sources such as social media \cite{mcneill_estimating_2017} prove inaccurate for transportation demand analysis \cite{camargo_diagnosing_2019}. Our empirical framework highlights an important finding that using digital ridership data we are able to extract a set of complete and accurate microscopic demand profiles which are also determinants of macroscopic urban structures in cities.

Though the data reveals a lot of information about the temporal and spatial profiles of traffic in a city, there are certain limiting factors to consider for future research. The model does not achieve high accuracy for predicting traffic (see Supplementary Note 4 for details). Even though our work does not focus on the accuracy of prediction, the data-driven method is able to highlight candidate station locations which suffer from poor ridership (both under and over represented), which in turn could be indicative of an unsatisfactory transport service that is a good case for improvement. Thus, the spatiotemporal geography we have presented is an important framework for assessing spatial use of a city with respect to its transportation infrastructure. Larger cities are increasingly interested in providing safe and secure travel options for nighttime workers and preserving or enhancing their nightlife as a cultural amenity and source of economic revenue. A report published by the Greater London Authority \cite{noauthor_london_2018} notes that fully a third of London workers work evening and nights, and two-thirds are actively engaging in nighttime activities. In addition, citizens may have very different expectations about how their districts should be used across time \cite{pinkster_when_2017} and cities are acknowledging those needs \cite{noauthor_london_2018}. Our results indeed confirm multiple uses of space over time and highlight the very specific districts where different kinds of activities occur, or might be enhanced with appropriate intervention. Our work can be used as a methodology for analysing and repurposing transport data for studies of cohesion, safety and growth.

Without requiring detailed origin-destination trajectories, which has become a common tool in urban studies, important information about urban structures could very well prompt studies in the direction of sustainable transit-oriented development \cite{zhang_network_2019,papa_accessibility_2015} and their potential negative impacts on displacing communities \cite{dawkins_transit-induced_2016}, promoting urbanisation \cite{kasraian_impact_2019} and exacerbating sprawl \cite{Bertaud2003,dieleman_compact_2004}. Our quantitative analysis of transportation demand has the potential to initiate new developments in extracting precise micro-scale \gls{od} matrices useful for urban planning on a city level by studying distributions of amenities around stations and correlating their use with generalised demand profiles. Since our method is generative, given mixed use of space, new stations can be planned or old ones re-designed, to match traffic demand for new technology hubs, social housing neighbourhoods or secondary or tertiary centres of tourism. Our method can be straightforwardly expanded to other transit networks, including multi-modal systems, and can therefore become a critical tool for urban and transit planners.

\section*{Methods} \label{sec:methods}
\subsection*{Transport Data}
We use the London Underground Passenger Count dataset as a proxy for ridership, which is provided freely by \gls{tfl} \cite{noauthor_london_nodate}. The dataset describes the average number of \emph{entrances} and \emph{exits} to and from each station in the Underground Network, represented as a time series spanning $24$ hours. The time series is aggregated at $15$-minute intervals, resulting in $96$ data points per station for a total of $264$ stations. We remove instances of erroneous or zero counts for every station between the times of $02:00am$ and $05:00am$ where the \gls{pt} network is not in function. This represents an average of all days in the month of November $2017$, separated into weekdays and weekends. We carry out our analysis on only the weekday data. The data description provided by \gls{tfl} claims that November $2017$ illustrates a typical sample of winter travelling behaviour in the year and has been adjusted for any disruptions in the Underground service (such as related to weather, malfunctions and accidents or community events). For details on sources of error, see Supplementary Note 4.

\subsection{Gaussian Mixture Model}
To identify the subpopulations of demand profiles in the ridership data, we formulate a \gls{gmm}. \gls{gmm}s are formed so subpopulations can be automatically learned from a large dataset without annotating any data points in advance with user-defined labels. Thus, a formulation of this type constitutes a class of unsupervised learning algorithms. The foundation of these models is built upon a mixture of several normal distributions. In the case of a passenger count dataset of a public transit system, our assumption is that the underlying distribution of the overall traffic at a station per day follows the sum of multiple scaled normal distributions \ref{fig:fig2} with their means at different times representing local minima. Translating this mixture onto a standard \gls{gmm} means that each distribution learned from the model represents one specific temporal demand profile. It is useful to observe than a data point in any one normal distribution does not necessarily classify that person entering a station as belonging to a specific demand type. A \gls{gmm} is probabilistic in nature that associates a likelihood to each data point of belonging to a specific normal distribution.

A \gls{gmm} is characterised by three parameters: individual mixture weights $\phi_k$, mixture means $\mu_k$ and variances $\sigma_k$ for all mixtures $C_k$. For our case of \gls{tfl} data, let the number of mixtures be $k=6$ (see Supplementary Figure 6 for effects of variations in $k$). Given our input data in the form of a time-series and the number of demand profiles (mixtures) we specify, the formulated \gls{gmm} first estimates a-posteriori the unknown parameters ($\phi_k, \mu_k, \sigma_k$) using an expectation maximisation algorithm \cite{dempster_maximum_1977}.

The probability distribution of the data $x$ in a \gls{gmm} is given by,

\begin{equation}
p(x) = \sum_{k=1}^{K} \phi_k \: \mathcal{N}(x \; \vert \; \mu_k,\,\sigma_k),
\end{equation}

\begin{equation}
\mathcal{N}(x \; \vert \; \mu_k,\,\sigma_k) = \frac{1}{\sigma_k \sqrt{2\pi}} \exp \left(
{-\frac{(x-\mu_k)^2}
{2 \sigma_k^{2}}}
\right), and
\end{equation}

\begin{equation}\label{eq:normalise}
\sum_{k=1}^{K} \phi_k = 1.
\end{equation}

Equation \ref{eq:normalise} shows that each mixture $C_k$ is marked by its associated weight such that the total probability distribution normalises to $1$.

\subsection{Generation of individual station traffic}
Using the Gaussian mixture model formulation and estimating the parameters for the cumulative set of observations in the data we are able to generate individual station traffic. The process of estimating the densities at every station involves two steps. First, we sample the normal mixture using the distribution of each identified demand profile $p(C_k) = \phi_k$. Second, we sample each data point belonging to the station from the distribution of mixture $C_k$ using $p(x \; \vert \; C_k) = \mathcal{N}(x \; \vert \; \mu_k,\,\sigma_k)$. Though we estimate densities at each station for observations that belong to the dataset, traffic for a new station that records out-of-sample observations could also be estimated using the parameters for our formulation.

\subsection{Clustering stations}
As the time series data of passenger counts is aggregated by $15$-minute intervals, we implicitly map the generated station traffic into a matrix where each row can be interpreted as station traffic over different times in a day. $P_i(t_j)$ is the measure of use of station $i$ at time-step $t_j$ (Table \ref{table:featureschema}). To curb the skewing effects of larger stations that also witness incoming and outgoing traffic to and from other cities, we normalise the passenger counts such that $\sum_{t} P_i(t) = 1 \; \forall$ stations $i$. Columns showing traffic over all stations at every time-step $t_j$ have long-tail distributions and each station traffic vector $[t_1,t_2,t_3,...,t_n]$ is a multi-modal distribution. Given this description, we formulate a multivariate \gls{gmm} for clustering stations into \emph{six} characteristic station types.

\begin{table}[!ht]
\centering
\caption{\textbf{Data schema for the feature matrix used for clustering.} Each station has $n$ features $t_j$ where every column (feature) represents the passenger traffic count $\vec{P}(t_j)$ at each time interval of $15$ minutes. As an example, $P_3(t_3)$ is the passenger traffic entering station $3$ at time interval $05:30am$ - $05:45am$, generated using our modelling approach.}
\begin{tabular}{ |c|c|c|c|c|c|  }
 \hline
 \multicolumn{6}{|c|}{Feature Matrix} \\
 \hline
 Station & $t_1$ & $t_2$ & $t_3$ & $t_{...}$ & $t_n$  \\
 \hline
 1 &. &. &. &. &. \\
 2 &. &. &. &. &. \\
 3 &. &. & $P_3(t_3)$ &. &. \\
 ... &. &. &. &. &. \\
 m &. &. &. &. &. \\
 \hline
\end{tabular}
\label{table:featureschema}
\end{table}

To model the multivariate \gls{gmm} case, we use the formulation,

\begin{equation}
p(\vec{x}) = \sum_{k=1}^{K} \phi_k \: \mathcal{N}(\vec{x} \; \vert \; \vec{\mu}_k,\, \Sigma_k),
\end{equation}

\begin{equation}\label{eq:multipdf}
\mathcal{N}(\vec{x} \; \vert \; \vec{\mu}_k,\, \Sigma_k) = \frac{1}{\sigma_k \sqrt{2\pi}} \exp \left(
{-\frac{(\vec{x}-\vec{\mu}_k)^2}
{2 \sigma_k^{2}}}
\right),
\end{equation}
where Eq.~\ref{eq:multipdf} is the probability density function of the multivariate normal distribution, $\vec{\mu}_k$ represents the means and $\Sigma_k$ the covariance matrices \cite{eirola_gaussian_2013}.

To cluster station traffic that follow similar trends over a day, we utilise the expectation step of the expectation-maximisation algorithm \cite{dempster_maximum_1977} which forms the basis of a \gls{gmm}. Using the estimated model parameters for the multivariate distributions, we find the likelihood that a station traffic pattern ($\vec{x}$) belongs to a mixture $C_k$ by calculating,

\begin{equation}
p(C_k \; \vert \; \vec{x}) = \frac
{\phi_k \: \mathcal{N}(\vec{x} \; \vert \; \vec{\mu}_k,\, \Sigma_k)}
{\sum_{k=1}^{K} \phi_k \: \mathcal{N}(\vec{x} \; \vert \; \vec{\mu}_k,\, \Sigma_k)}.
\end{equation}

\section{Data Availability}
All datasets that support the findings of this study are publicly available (as cited in the references) and a version used in the study can be collected, requested or directly downloaded from the following link: https://github.com/mikhailsirenko/spacetimegeo

\section{Code Availability}
The GMM representation together with all the necessary functions for running the model are available in python at the following link on github: https://github.com/mikhailsirenko/spacetimegeo

\bibliographystyle{unsrt}

\section{Author contributions}
All authors designed the study. TV, IK, MS, SC and NA evaluated the data. TV, MS and SC developed the model. All authors analyzed the results and wrote the manuscript.

\section{Competing interests}
The authors declare no competing interests.

\end{document}